\documentclass[manuscript,sigconf]{acmart}

\AtBeginDocument{%
  }

\setcopyright{none}
\renewcommand\footnotetextcopyrightpermission[1]{}
\acmConference[CSCW Companion '25]{Make sure to enter the correct
  conference title from your rights confirmation emai}{October 18--22,
  2025}{Bergen, NO}
\acmISBN{978-1-4503-XXXX-X/18/06}
\usepackage{listings}

\usepackage{xspace}
\newcommand{\whisper}[0]{\texttt{whisper-1}\xspace}
\newcommand{\fouromini}[0]{\texttt{gpt-4o}\xspace}
\newcommand{\ellabs}[0]{\texttt{elevenlabs-tts}\xspace}

\usepackage{tcolorbox}
\definecolor{accent}{HTML}{DC1E61}

\begin{document}

\title{PAL: Designing Conversational Agents as Scalable, Cooperative Patient Simulators for Palliative‑Care Training}

% Here's an alternative title:
% PAL: Transforming Palliative Care Communication Training with Empathetic Chatbots


\author{Neil K. R. Sehgal}
\authornote{These authors contributed equally to this work.}
\email{nsehgal@seas.upenn.edu}
\affiliation{%
  \institution{University of Pennsylvania}
  \city{Philadelphia}
  \state{Pennsylvania}
  \country{USA}
}

\author{Hita Kambhamettu}
\authornotemark[1] % Refers to the previous \authornote
\email{hitakam@seas.upenn.edu}
\affiliation{%
  \institution{University of Pennsylvania}
  \city{Philadelphia}
  \state{Pennsylvania}
  \country{USA}
}
\author{Allen Chang}
\authornotemark[1] % Refers to the previous \authornote
\email{cylumn@seas.upenn.edu}
\affiliation{%
  \institution{University of Pennsylvania}
  \city{Philadelphia}
  \state{Pennsylvania}
  \country{USA}
}

\author{Andrew Zhu}
\authornotemark[1] % Refers to the previous \authornote
\email{andrz@seas.upenn.edu}
\affiliation{%
  \institution{University of Pennsylvania}
  \city{Philadelphia}
  \state{Pennsylvania}
  \country{USA}
}

\author{Lyle Ungar}
\email{ungar@cis.upenn.edu}
\affiliation{%
  \institution{University of Pennsylvania}
  \city{Philadelphia}
  \state{Pennsylvania}
  \country{USA}
}

\author{Sharath Chandra Guntuku}
\email{sharathg@seas.upenn.edu}
\affiliation{%
  \institution{University of Pennsylvania}
  \city{Philadelphia}
  \state{Pennsylvania}
  \country{USA}
}

\renewcommand{\shortauthors}{Sehgal et al.}

\begin{abstract}

Effective communication in serious illness and palliative care is essential but often under-taught due to limited access to training resources like standardized patients. We present PAL (Palliative Assisted Learning-bot), a conversational system that simulates emotionally nuanced patient interactions and delivers structured feedback grounded in an existing empathy-based framework. PAL supports  text and voice modalities and is designed to scaffold clinical skill-building through repeated, low-cost practice. Through a mixed-methods study with 17 U.S. medical trainees and clinicians, we explore user engagement with PAL, evaluate usability, and examine design tensions around modalities, emotional realism, and feedback delivery. Participants found PAL helpful for reflection and skill refinement, though some noted limitations in emotional authenticity and the adaptability of feedback. We contribute: (1) empirical evidence that large language models  can support palliative communication training; (2) design insights for modality-aware, emotionally sensitive simulation tools; and (3) implications for systems that support emotional labor, cooperative learning, and AI-augmented training in high-stakes care settings.

\end{abstract}





\maketitle

\section{Introduction}
Communication during serious illness is a high-stakes form of clinical collaboration often undertaught in medical training. Palliative care conversations -- such as delivering a prognosis, discussing end-of-life preferences, or responding to patient and family emotions -- require empathy, clarity, and adaptability. Communication missteps can result in misunderstandings, decreased patient satisfaction, and potential harm \cite{Hassan2018}. Yet, existing training approaches are hard to scale, leading many clinicians to enter practice underprepared for these emotionally complex interactions \cite{fraser2001senior, sulmasy2008us, sullivan2004end}.

Training physicians for these conversations  often involves Standardized Patients (SPs), trained actors who simulate patient scenarios and provide feedback.
However, SPs can be difficult to recruit, expensive to train, and subject to scheduling constraints for both trainees and facilitators \cite{NewlinCanzone2013, Hart2016, Brenner2009, Nagpal2021}. 
Further, SPs can be inconsistent in performance and often report difficulties in giving proper feedback \cite{Hart2016, abe2011nationwide}.
For example, the cost to secure SPs and SP trainers for about 15 clinicians through our academic institution's health system is around \$6,000, with an additional \$1,000 for a faculty coordinator.
These limitations may restrict the accessibility and effectiveness of SP-based training for institutions with limited resources.

Alternatively, rules-based virtual patient simulators can cheaply simulate patient scenarios for basic skills training, such as gathering patient history.
However, most existing commercial simulators lack the ability to react emotionally, provide implicit feedback, or convincingly mimic the nuances of human interaction, which is crucial in palliative care scenarios where patients' emotions are vulnerable \cite{Kang2024}.

Human–computer interaction (HCI) research has begun to address this gap, exploring how conversational agents, intelligent feedback systems, and cooperative technologies can augment communication training in sensitive domains. In this work, we introduce PAL (Palliative Assisted Learning-bot)—a web-based chatbot that simulates emotionally nuanced patient interactions and delivers structured, scenario-specific feedback. PAL supports both text and voice modalities and is powered by large language models (LLMs), enabling trainees to rehearse difficult conversations on demand at low cost.

Through a mixed-methods study with 17 U.S. clinicians and trainees, we evaluate PAL’s usability, modality design, and feedback mechanisms. We find that PAL enables skill-building across experience levels, while surfacing design tensions around realism, empathy frameworks, and adaptive training. This work contributes empirical evidence for LLM-driven simulation in palliative care, a design framework for chatbot-mediated training, and implications for systems that support emotional labor and clinical collaboration at scale.

\newpage

\section{Related Work}
This work builds on a growing body of HCI and CSCW research focused on designing for serious illness and end-of-life care. Prior work has emphasized the emotional, social, and infrastructural complexity of palliative communication.  Ferguson et al. emphasize the multifaceted role of technology in supporting comfort and communication during end-of-life care, arguing for designs that reflect the evolving needs of caregivers and patients \cite{ferguson2014craving}. Similarly, Chen et al. examine how caregiving impacts emotional well-being and advocate for sociotechnical systems that recognize caregiver needs, not just those of patients \cite{chen2013caring}. These works underscore the collaborative, multi-stakeholder nature of serious illness care and the need for HCI systems that navigate emotional nuance and logistical complexity.

Recent advances in LLMs and AI-driven simulation have opened new avenues for scalable, immersive medical training. For instance, the MedSimAI system uses LLM-powered simulated patients with structured feedback to support clinical communication skill development, with a focus on routine clinical interactions such as patient history taking and diagnostic interviews \cite{hicke2025medsimai}.  Similarly, other work has used avatar-based patients for nursing communication training and chatbot-based modules to improve exam scores on standardized medical interviewing exams \cite{shorey2019virtual, yamamoto2024enhancing}. While these studies demonstrate the viability of conversational agents in clinical education,  past work often sidesteps emotionally fraught situations such as serious illness diagnosis, where clinicians must navigate nuanced patient emotions, distress, and end-of-life considerations. This represents a key gap in current simulation design.

A parallel line of work explores emotionally responsive AI feedback in educational contexts. Feedback is central to communication training, yet prior work in health education has shown that receiving negative feedback can evoke negative reactions and limit reflection \cite{sargeant2008understanding}. AI-based systems offer opportunities for adaptive, emotionally-sensitive just-in-time feedback that supports skill development without fully replacing human mentorship. For instance, Alsaiari et al. demonstrate that generative AI feedback enriched with emotional tone --- such as praise or empathy --- can mitigate negative emotional reactions like anger and improve student perceptions of feedback utility, even if it does not significantly change work quality or engagement levels \cite{alsaiari2024emotionally}. Our work extends this line of inquiry by exploring how LLM-based chatbots might offer reflective, emotionally aware feedback for skill-building in palliative care communication.

PAL contributes a novel integration of clinical empathy pedagogy and LLM-based simulation. First, it is explicitly designed for palliative and end-of-life communication, a domain underrepresented in prior simulation work despite its emotional complexity and clinical urgency. Second, PAL is grounded in a validated empathy framework (NURSE), translating abstract communication goals into concrete, interpretable feedback. This positions PAL as a unique bridge between emotionally intelligent coaching systems and the demands of high-stakes clinical care.

\section{System Design}
PAL is a web-based chatbot that simulates emotionally nuanced patient interactions to support palliative care communication training. It supports both text and voice modalities and delivers post-session feedback grounded in the NURSE empathy framework (Naming, Understanding, Respecting, Supporting, Exploring). Users engage in simulated serious illness conversations, with PAL adapting its tone and emotional responses based on clinician input.

\textbf{Patient Simulation and Feedback
} Patient personas are adapted from real standardized patient (SP) training scenarios and include background, clinical setting, and emotional trajectory (e.g., denial, grief). The underlying chatbot—powered by GPT-4o—generates context-sensitive replies that reflect emotional realism. After each session, PAL provides structured, scenario-specific feedback: missed opportunities for empathy are highlighted using direct quotes from the transcript, along with recommended phrasing improvements. Feedback is produced from GPT-4o's analysis of the transcript, guided by prompt-engineered instructions and grounded in the NURSE medical communication framework \cite{NURSE}.

The NURSE framework articulates empathy through five core strategies: Naming, Understanding, Respecting, Supporting, and Exploring.
This framework is designed to address emotional cues effectively, fostering deeper patient-clinician understanding. Each actionable recommendation from PAL is structured as the scenario, the current approach (a quote from the transcript), and an improvement suggestion based on the NURSE framework.
For example, when a clinician omits an acknowledgment of an emotion, the feedback suggests integrating a statement such as, ``I know this must be difficult to hear,'' to provide emotional validation.
The full feedback generation prompt is given in the appendix. 

\textbf{Interface and Technical Implementation
} PAL is built using FastAPI and Vue.js, allowing two-way streaming of text and voice interaction. In text mode, users type responses and receive real-time replies with embedded emotional cues (e.g. italicized annotations indicating the patient is pausing, crying, or showing hesitation). In voice mode, users press-to-speak via microphone input; PAL replies via text-to-speech (TTS) audio only. All interactions conclude with on-screen textual feedback summaries. The system uses Whisper for voice-to-text transcription and ElevenLabs for TTS generation. For implementation specifics and an image of the interface, please refer to the appendix.

\section{Methods}
We conducted a mixed-methods usability study to evaluate PAL’s design and training potential, focusing on three research questions:

\textit{RQ1: How do clinicians engage with PAL across voice and text modalities?}

\textit{RQ2: What are the advantages and limitations of chatbot-mediated training?}

\textit{RQ3: How can such systems be improved for serious illness communication?}

\textbf{Participants} We recruited 17 U.S.-based medical professionals across training levels—medical students (n=5), residents (n=5), and attending physicians (n=7)—via institutional email lists and snowball sampling. Participants had varying exposure to formal palliative care communication training.

\textbf{Interview Procedure}
Each participant completed a 60-minute remote session via Zoom, structured into three phases: (1) a 10-minute formative interview on their training experience; (2) two 20-minute hands-on interactions with PAL, split between text and voice modalities (order counterbalanced); and (3) post-interaction reflection and feedback via an open-ended interview and a short survey. Participants completed one full scenario per modality. See appendix for full interview procedure and interview questions. 

\textbf{Data Collection and Analysis}
All interviews were recorded with participant consent and transcribed by Zoom for analysis. A two-pass thematic analysis was conducted: the first author performed open coding to identify emergent themes; a second author refined and reviewed the codebook. Disagreements were resolved collaboratively. We prioritized thematic validity over inter-rater reliability to preserve contextual nuance \cite{mcdonald2019reliability, kambhamettu2024explainable}. Quantitative usability ratings (ease of use, realism, learning value) were averaged across participants. The study protocol was approved by the institutional IRB; informed consent was obtained. Total system cost for the 17 interviews, including 3 months of server and domain hosting, as well as API usage for \whisper, \fouromini, and \ellabs was \$32.59.

\section{Findings}
Our findings are organized into four themes: (1) challenges in learning serious illness communication, (2) participant experiences with PAL, (3) modality-specific insights, and (4) recommendations to improve the system.

\subsection{Challenges in Learning}
Despite the growing inclusion of serious illness communication in medical curricula \cite{fitzpatrick2017palliative}, many participants—especially practicing physicians—reported little to no formal training (n=4). Those that did have formal training still felt that there were significant gaps in their ability to communicate to patients with serious illnesses, because they were unable to anticipate tricky patient questions (n=3) and do not have access to in-depth training opportunities (n=3). Instead, many described learning "on the job" through observation and trial and error. "The vast majority of learning comes from how other people break bad news," explained one resident (P9). "The different attending doctors that we have, they all have very different style...I have learned a lot from watching other people" (P10). One physician recalled being asked as a medical student to deliver end-of-life information alone, with no prior preparation: "It was really scary... I just grabbed the hospital chaplain...and said, could you go with me?" (P2).

Even when formal instruction was available, participants described it as overly high-level or lecture-based, with limited practice opportunities. "[They] give us a PowerPoint about how to deliver bad news...without practicing it...it’s way less meaningful" (P7). Across experience levels, participants emphasized the importance of developing a personal communication style, shaped more by imitation and improvisation than structured learning. Yet, emotionally charged questions—such as "How long do I have?"—were seen as persistently difficult. "It’s hard to answer questions when people want exact predictions," one participant noted (P2). Burnout and clinical time pressures further constrained reflection and growth, with participants admitting that ‘softer’ skills like communication often took a backseat to technical learning. "Burnout is super high ... [it] discourages people from ... continuing to think about how they can get better at what they’re doing," explained P10.

\subsection{Strengths of the Chatbot}
Participants responded positively to PAL and reported strong usability and training value. On a 1–5 scale, participants rated PAL highly for ease of use (4.5), feedback utility (4.1), conversation relevance (4.5), difficulty (4.3), and skill learning (4.0). They appreciated the structured feedback and its ability to prompt reflection. "I was impressed with the level of detail," said one student (P6). "It was able to look back at the conversation in a really detailed way...that was super helpful" (P6).

Several participants described noticeable learning effects, reporting improvement between sessions. "I gave a little too much information in the first session... so [in the second] I gave the patient more time to process" (P7). Others reflected on skill gaps with openness: "I feel like I’m less of a doctor... I need lots of help, that’s what I know" (P1). P11 echoed, "My lack of clinical experience is showing." However, both expressed positive sentiments when sharing these thoughts, and overall, the bot was seen as a supportive, low-pressure training environment.

\subsection{Modality-Specific Experiences}
Most participants preferred the voice modality for its realism (n=12). "The voice mode felt more realistic because that’s how we actually communicate" (P16). Others described it as "more like a real patient interaction" (P10), helping them practice verbal delivery under pressure. At the same time, many found the text modality helpful for crafting responses in a more thoughtful manner. "I felt less nervous doing it over text. It’s easier to correct yourself" (P16). Some (n=4) recommended beginning with text before progressing to voice to build confidence: "It would be helpful to practice with both...start with the text chatbot, then move on to the voice one" (P7).

However, both modalities posed challenges. Several participants found typing unnatural for simulating real conversations. "It feels so unnatural to be typing. You weren’t leaving any silences or any pauses" (P9). Others  noted that the push-to-talk design in the voice interface disrupted conversational flow. "It was awkward pressing the button to talk and then pressing again when I'm done" (P2). P8 likened the voice input to dictation, reducing the sense of immersion.

\subsection{Recommendations for Improvement}
\textbf{Increase Emotional Authenticity:} Participants suggested subtle additions to increase emotional authenticity—such as pauses, emotionally expressive phrases, and nuanced tone variation. "When you tell someone they have cancer, they take a pause," one clinician noted (P16). They also recommended more expressive verbal cues (e.g., "I can’t believe this is happening") and more emotional language to reflect real patient reactions.

\textbf{Diversify Patient Training Cases:} Many participants wanted broader training scenarios beyond cancer, such as dementia, heart failure, and family-centered conversations. "Everybody knows about cancer... 
it’s a lot harder to have a conversation when the person has dementia...there's other conversations you have to have...[like] taking away their driving privileges" (P1). Others proposed multi-day or sequential case formats to simulate evolving care conversations over time.

\textbf{Flexible Feedback Delivery:} While most participants valued PAL’s structured feedback, several noted that the NURSE framework felt too rigid or misaligned with their personal style. "The feedback was helpful, but sometimes the NURSE framework isn’t the most useful for real-life conversations" (P6). More experienced physicians were occasionally resistant, describing the feedback as patronizing or overly prescriptive. "This is like the classic this is gonna help other people but it won't help me" (P8). Suggestions included incorporating more positive reinforcement, adding real-time coaching ("Maybe you want to try it this way"), and escalating the negative emotion of the chatbot's responses when communication from the clinician was poor. A few participants expressed interest in tracking progress across sessions through performance scores or trend visualizations.

\textbf{Support Hybrid Training:} Finally, many participants emphasized that PAL should be positioned as a complement—not a replacement—for live training. "I see this as being great practice before you go in with a live standardized patient" (P15). Others saw value in offloading portions of training to the chatbot to free up in-person time for deeper reflection. A hybrid approach—leveraging the scalability of AI with the richness of human feedback—was seen as most effective for developing nuanced, empathetic communication skills.

\section{Discussion}

This study demonstrates how conversational agents can support communication training in emotionally demanding clinical work, such as palliative care. By enabling asynchronous, on-demand practice with emotionally responsive patient simulations, PAL offers an alternative to traditional SP methods that are costly, logistically constrained, and difficult to scale. Our findings suggest that AI-driven systems can serve as collaborative learning partners, not only facilitating rehearsal, but prompting self-reflection and adaptation in communication style.

This study has several limitations. First, our sample size (n=17) was relatively small and limited to clinicians and trainees in the U.S., which may affect the generalizability of findings across cultural or institutional contexts. Second, while PAL's simulated interactions elicited meaningful reflections, we did not assess long-term skill retention or transfer to real clinical practice. Third, the emotional authenticity of AI-generated responses, though rated positively by many participants, still falls short of the complexity found in real human encounters, particularly in high-stakes, multimodal communication. Finally, while we focused on feedback delivered through the NURSE framework, future work should explore adaptive feedback models that align with varying communication styles and cultural expectations.

A key contribution of this work lies in examining how AI systems might scaffold emotion work in cooperative clinical practice. Serious illness communication often involves not just medical information transfer, but empathetic negotiation, emotion recognition, and relational alignment with patients and families \cite{ferguson2014craving, back2009mastering}. Participants in our study highlighted both the utility and the limitations of structured frameworks like NURSE in guiding such work. While some found the feedback actionable, others felt the feedback was rigid, potentially conflicting with previously developed personal or culturally grounded communication styles. This tension underscores a central design challenge: How to design flexible scaffolds that support, rather than prescribe, professional discretion and affective labor.

The dual-modality design of PAL, text and voice, also illuminates how tool-mediated interaction can support different stages of learning. Text-based interactions enabled reflection and iterative practice, while voice interactions introduced real-time pressure and social realism. Participants often suggested a progression from text to voice.

Participants also expressed interest in hybrid integration, using PAL as a precursor or supplement to SP sessions or peer learning. This suggests a role for conversational agents in blended training, where AI augments human-led education, showing how LLM-driven agents fits into institutional workflows without replacing critical human elements.

Finally, the desire for more personalized, dynamic feedback--such as real-time suggestions or longitudinal tracking--points to opportunities for AI systems that act as reflective collaborators in skill development. At the same time, participants noted that emotionally vulnerable tasks require sensitivity in feedback delivery. These observations highlight the need for future systems to balance automation with affective attunement, ensuring that feedback mechanisms preserve learner agency and emotional safety.

\section{Conclusion}

This work demonstrates the promise of PAL in addressing the need for scalable, cost-effective palliative care communication training.
By simulating emotionally sensitive interactions and offering structured feedback, PAL helps clinicians improve communication skills in a low-risk environment.
Our findings highlight chatbot strengths including flexible text and voice modalities for complementary skill building. We also assess current limitations and outline a path toward improved training guided by clinician recommendations.
With continued refinement and integration into broader curricula, PAL can complement traditional methods, enhancing the accessibility and quality of communication education in palliative care.

\bibliographystyle{ACM-Reference-Format}
\bibliography{sample-base}

\clearpage
\onecolumn
\appendix
\title{PAL: Transforming Palliative Care Communication Training with Empathetic Chatbots}

% Here's an alternative title:
% PAL: Transforming Palliative Care Communication Training with Empathetic Chatbots

\author{Neil}
\authornote{These authors contributed equally to this work.}
\affiliation{%
  \institution{University of Pennsylvania}
  \city{Philadelphia}
  \state{PA}
  \country{USA}
}

\author{Hita}
\authornotemark[1] % Refers to the previous \authornote
\affiliation{%
  \institution{University of Pennsylvania}
  \city{Philadelphia}
  \state{PA}
  \country{USA}
}

\author{Allen}
\authornotemark[1] % Refers to the previous \authornote
\affiliation{%
  \institution{University of Pennsylvania}
  \city{Philadelphia}
  \state{PA}
  \country{USA}
}

\author{Andrew}
\authornotemark[1] % Refers to the previous \authornote
\affiliation{%
  \institution{University of Pennsylvania}
  \city{Philadelphia}
  \state{PA}
  \country{USA}
}

\renewcommand{\shortauthors}{Sehgal et al.}


\appendix

\section{Detailed System Design}

\begin{figure}[H]
\centering
\includegraphics[width=\textwidth]{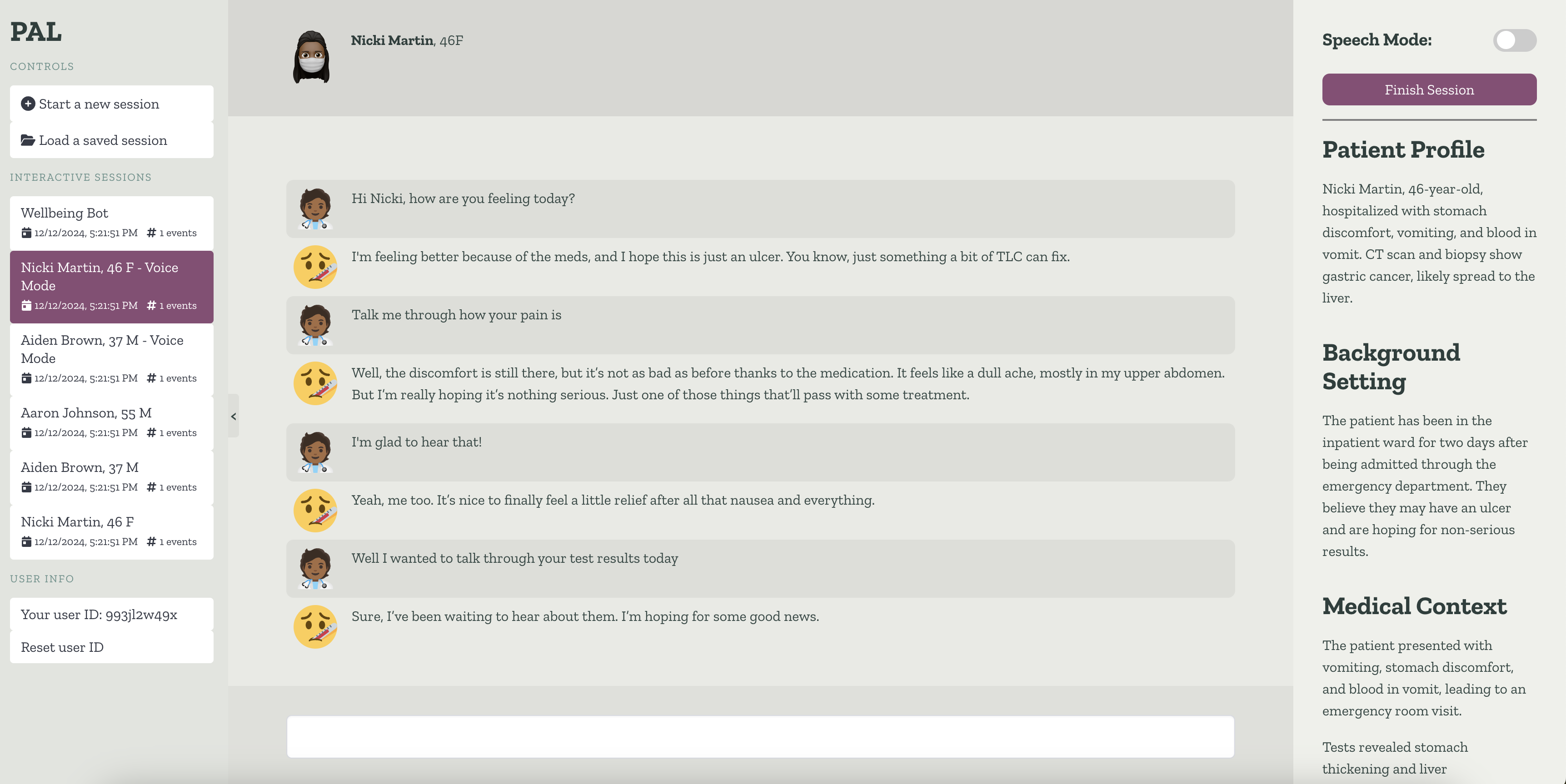}
\caption{The PAL system: in the main display a provider can interact with the patient bot. On the right side panel, there is a speech modal key, a button to finish the session which triggers feedback report generation, and a patient profile which details information about the patient and their reason for visiting. On the left side panel are different patient personas.}
\end{figure}

\subsection{Patient Agent Design}
PAL's patient agent profiles are based on existing standardized patient cases, ranging from 2 to 12 pages. Each case includes:
\begin{itemize}
    \item \textbf{Purpose of the Case}: Specifies learning objectives, such as delivering bad news.
    \item \textbf{Persona of the Patient}: Defines emotional disposition, attitudes, and behavior.
    \item \textbf{Past Medical History}: Summarizes diagnoses, treatments, and symptoms.
    \item \textbf{Social History}: Outlines family dynamics, profession, and cultural background.
    \item \textbf{Setting}: Describes the clinical context (e.g., hospital ward).
    \item \textbf{Stages of the Interaction}: Guides responses through different dialogue phases.
\end{itemize}

The prompts are derived from these cases to capture emotional, social, and medical backgrounds. Specific features include:
\begin{itemize}
    \item Simulating emotionally nuanced responses tailored to user communication.
    \item Mirroring emotional progression, such as optimism giving way to denial or acceptance.
    \item Adapting responses dynamically to empathetic or non-empathetic cues.
\end{itemize}

As the prompts are derived from training materials from an existing program we are unnable to share them in full.

\subsection{User Interface}
PAL's web application uses Vue.js and incorporates real-time streaming for both text- and speech-based interactions. Users are assigned unique IDs for interaction logging and feedback purposes. The interface consists of:

\paragraph{Control Panel}
Users can start new sessions or load previous ones. Sessions are listed chronologically with metadata such as patient age, gender, and profile image.

\paragraph{Interactive Session}
\begin{itemize}
    \item \textbf{Text-Based Interactions:} A text input area allows users to type their statements, with responses streamed in real time to emulate natural conversation.
    \item \textbf{Speech-Based Interactions:} Users press a button to record their speech, which is transcribed using \whisper, processed by \fouromini, and synthesized into audio responses via \ellabs. Audio responses are streamed without transcripts, mimicking a phone call.
\end{itemize}

\paragraph{Patient Information Panel}
The panel provides demographic and medical details, such as age, symptoms, diagnoses, and the patient's understanding of their condition. This ensures users have context for an empathetic and informed clinical encounter.

\paragraph{Session Termination and Feedback}
The ``Finish Session'' button ends the session and sends an automatic feedback summary to the user.

\section{Detailed Interview Procedure}

\subsection{Phase 1: Formative Interview (10 minutes)}
Participants answered open-ended questions to contextualize their experiences with palliative care communication and training. Example questions included:
\begin{itemize}
    \item ``How has your experience been with helping people talk through palliative care options?''
    \item ``What methods have you used to learn these skills? For example, interacting with standardized patients, attending seminars, or talking to peers?''
    \item ``What challenges do you encounter with your current learning strategies?''
    \item ``What kinds of support or resources would you find most helpful in training?''
\end{itemize}

\subsection{Phase 2: Interaction with Chatbot (30 minutes)}
Participants interacted with two chatbot modalities: text-based and voice-based.
Each interaction lasted up to 15 minutes, followed by feedback.

\subsubsection{Text Modality Interaction}
Participants used a text-only chatbot interface to simulate conversations with a virtual patient.
They were instructed to:
\begin{itemize}
    \item ``Message as if this bot is a patient.''
    \item ``Think out loud about what aspects of this process work well and what difficulties arise.''
\end{itemize}

\subsubsection{Voice Modality Interaction}
Participants engaged with a voice-only chatbot using a push-to-talk input method.
Instructions included:
\begin{itemize}
    \item ``Speak as if this bot is a patient.''
    \item ``Think out loud about what aspects of this process work well and what difficulties arise.''
\end{itemize}

\paragraph{Follow-Up Questions}
After each interaction, participants completed a Google Form, which included a 5-point Likert scale evaluation of the chatbot based on:
\begin{enumerate}
    \item Ease of use
    \item Usefulness of feedback
    \item Relevance of the conversation
    \item Appropriate difficulty of the conversation
    \item Skill acquisition applicability
\end{enumerate}

Participants were then asked a series of open-ended questions such as:
\begin{itemize}
    \item ``What was your least favorite part? Your most favorite part?''
    \item ``What was particularly easy? Particularly hard?''
    \item ``Did you feel like the feedback was sufficient?''
    \item ``What parts of the conversation were most useful?''
\end{itemize}

\subsection{Phase 3: Closing Questions (10 minutes)}
The session concluded with reflective questions designed to gather additional insights:
\begin{itemize}
    \item ``If you had a magical tool that could do everything you wanted to help you with this problem, what would it do?''
    \item ``Before we wrap up, are there any last burning comments you might want to share?''
\end{itemize}

\newpage

\section{Participant Backgrounds}

\begin{table}[!h]
\centering
\begin{tabular}{|c|l|c|}
\hline
\textbf{Participant ID} & \textbf{Role} & \textbf{Gender} \\ \hline
P1 & Primary Care MD & Woman \\ \hline
P2 & Nephrology MD & Man \\ \hline
P3 &  Primary Care MD & Woman \\ \hline
P4 & Medical Student & Man \\ \hline
P5 & Medical Student & Woman \\ \hline
P6 & Medical Student & Woman \\ \hline
P7 & Medical Student & Woman \\ \hline
P8 & Emergency Medicine MD & Man \\ \hline
P9 & OB-GYN Resident & Woman \\ \hline
P10 & OB-GYN Resident & Woman \\ \hline
P11 & Medical Student & Woman \\ \hline
P12 & Medical Student & Woman \\ \hline
P13 & Resident & Man \\ \hline
P14 & MD-PhD Student & Woman \\ \hline
P15 & Hospice and Palliative Care MD & Man \\ \hline
P16 & Palliative Care MD & Woman \\ \hline
P17 & Emergency Medicine MD & Woman \\ \hline
\end{tabular}
\caption{Participant Background}
\label{tab:participant_roles}
\end{table}

\section{Feedback System Prompt}

\lstset{
    basicstyle=\ttfamily\small,
    frame=single,
    breaklines=true,
    columns=fullflexible,
    numbers=left,
    numberstyle=\tiny,
    stepnumber=1,
    showspaces=false,
    showstringspaces=false,
    breakatwhitespace=true,
}

\begin{lstlisting}
Analyze a transcript from a doctor-patient encounter and provide actionable communication improvement advice for the doctor.

Consider effective communication tools such as NURSE statements, avoiding jargon, and preventing common learner hiccups. Your advice should be clear, specific, and include practical steps for improvement. Address emotional cues and provide suggestions to optimize patient understanding and support.

# Steps

- **Analyze Transcript:**
  - Identify moments where the doctor's communication can be improved.
  - Assess instances where the doctor faced emotionally driven reactions from the patient/family, and determine whether appropriate NURSE statements were used.
  
- **Identify Gaps and Give Actionable Feedback:**
  - Focus on common learner hiccups like skipping initial steps, unclear headlines, or neglecting to offer NURSE statements.
  - Offer suggestions that go beyond simple feedback, outlining specific ways the doctor can alter their phrasing or behavior.

- **Provide Emotional Support Guidance:**
  - When providing alternate suggestions, use examples that appropriately name emotions, offer understanding or respect, explore emotions, or provide emotional support.

- **Link Feedback to Techniques:**
  - Clearly link feedback to provided communication methods such as NURSE statements, offering enhanced implementation or corrections.

# Output Format

Provide feedback in a list format where:
- Each item contains **a specific scenario/moment** that could be improved.
- Each item includes **detailed suggestions** for what the doctor could say differently, and why this change is beneficial.
- Use **NURSE-related language** where applicable and avoid broad, unspecific comments.

Example Feedback Segment:
1. **Scenario**: Doctor introduces prognosis without assessing the emotion.
   - **Current Approach**: "The prognosis is not very good."
   - **Improvement Suggestion**: Add an understanding statement first. For instance, "I know this must be really hard to hear." This would help in validating the patient's feelings, allowing them space to process the news and feel understood.
  
2. **Scenario**: Doctor uses medical jargon.
   - **Current Approach**: "There's evidence of multisystem organ failure."
   - **Improvement Suggestion**: Replace jargon with simpler language. Try, "We're noticing that several of their important organs are starting to not work as they should." This ensures that the patient and their family can follow and understand the diagnosis clearly.

3. **Scenario**: Doctor doesn't explore the family member's concerns after giving troubling news.
   - **Current Approach**: "This must be really hard. But let's talk about next steps."
   - **Improvement Suggestion**: Instead, pause after acknowledging their concerns with an explore statement: "Can you tell me more about what's on your mind right now?" This gives an opportunity for the family member to voice their concerns and ensures they feel heard.

# Notes

- **Avoid Fake NURSE Statements**: Ensure the given NURSE statement is sincere, and allow adequate space for the patient to react.
- **Avoid Jargon or Vague Headline Information**: Deliver information directly with clear, patient-friendly language.
- **Adjust Emotional Attunement**: Pay attention to phrases like "I understand," which can come across as overconfident about a family's experience. Use phrases like "I can't imagine what you're feeling" to show empathy without overstepping.
\end{lstlisting}

\end{document}